\documentclass[a4paper]{article}  %>>> use this instead for A4 paper
\usepackage[]{authblk}
\usepackage[]{graphicx}

\title{A new algorithm for sky extraction for multi-fiber instrument} 

\author[1,2]{Rodrigues M.}  \author[1]{Flores H.} \author[1]{Puech, M.} \author[3]{Yang Y}  \author[1]{Royer. F}
 \affil[1]{GEPI -Observatoire de Paris, 5 place Jules Janssen, 92190 Meudon, France}
 \affil[2]{CENTRA - Instituto Superior Tecnico, Av. Rovisco Pais, 1049-001 Lisboa, Portugal}
  \affil[3]{National Astronomical Observatories, 20A Datun Road, 100012 Beijing, PR China}
\begin{document} 
  \maketitle 

%%%%%%%%%%%%%%%%%%%%%%%%%%%%%%%%%%%%%%%%%%%%%%%%%%%%%%%%%%%%% 
\begin{abstract}
We present a new method to subtract sky light from faint object
observations with fiber-fed spectrographs. The algorithm has been
developed in the framework of the phase A of OPTIMOS-EVE, an
optical-to-IR multi-object spectrograph for the future european
extremely large telescope (E-ELT). The new technique overcomes the
apparent limitation of fiber-fed instrument to recover with high
accuracy the sky contribution. The algorithm is based on the
reconstruction of the spatial fluctuations of the sky background (both
continuum and emission) and allows us to subtract the sky background
contribution in an FoV of $7\times7\,arcmin^2$ with an accuracy of 1\%
in the mono-fibers mode, and 0.3-0.4\% for integral-field-unit
observations.
\end{abstract}

%%%%%%%%%%%%%%%%%%%%%%%%%%%%%%%%%%%%%%%%%%%%%%%%%%%%%%%%%%%%%
\section{INTRODUCTION}
\label{sec:intro}  % \label{} allows reference to this section
\subsection{OPTIMOS-EVE}
\begin{figure}[!h]
\centering
\includegraphics[width=0.82\textwidth]{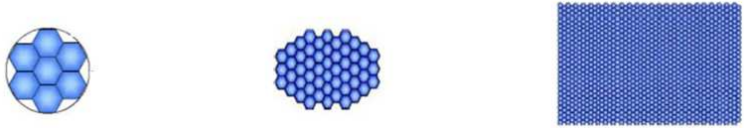}
\caption{{\small The three observational modes of OPTIMOS-EVE. From
    left to right: (1) the multi-object mode (MO).The MO LR mode will
    allow us to observed at low spectral resolution (LR, with
    R=5\,000) 240 objects with a total aperture on sky of 0.3". The
    medium and high resolution modes will allow us to observe 70 and
    40 objects within a total aperture of 0.9" and 0.72" and a
    spectral resolution of 15\,000 and 30\,000, respectively; (2) the
    medium Integral Field Unit (MI): 30 IFUs composed of 56 fibers
    covering a surface of 2"x3". The spatial sampling is 0.3" and the
    spectral resolution is R=5\,000; (3) Large integral field unit
    mode (LI). It is a large 15"x9" IFU composed of 1650 fibers. The
    spectral and spatial resolutions are the same that in the MI
    mode.}}
\label{EVE_mode}
\end{figure}

OPTIMOS-EVE (the Extreme Visual Explorer, PI: F. Hammer, GEPI and
co-PI Lex Kaper, UvA) is a fibre-fed, optical-to-infrared multi-object
spectrograph designed to explore the large field of view provided by
the E-ELT at seeing limited conditions. It will enable to observe
simultaneously multiple scientific targets in a 7' field of view with
on of the three fiber setups provided, see figure \ref{EVE_mode}. To
achieve the scientific goals of OPTIMOS-EVE, the sky background has to
be extracted with an accuracy $<$1\%.

\subsection{Sky subtraction in fiber-fed instrument}
Multi-fibre spectrographs are often suspected to suffer from a major
drawback: the low accuracy of the sky subtraction process. Such a
limitation is a critical issue for OPTIMOS-EVE, which aims to observe
very faint objects. Contrary to slit spectroscopy that samples sky
directly next to the object, the sky subtraction process in fiber
spectroscopy is more problematic because the sky contribbution cannot
be sampled at the immediate vicinity of the target. This comparison
has led to a commonly held view that accurate sky subtraction cannot
be achieved with a multi-object fibre spectrograph. Several cons have
been cited in litterature against accurate sky subtraction with
fiber-fed instruments, such as: spatial variations of the sky
background, spatial variations of the quantum efficiency of the
detector, variation of the fiber-to-fiber response (transmission,
instrument flexure), scattered light, cross-talk between fibers
\cite{1992MNRAS.257....1W}, \cite{1998ASPC..152...50W}. 

To overcome these difficulties, two standard observational procedures
have been implemented to improve the efficiency of the sky subtraction
process with fiber-fed instruments:
\begin{itemize}
\item \textbf{Simultaneous background:} Several fibers are dedicated
  to sample the sky contribution in the observed region. The number of
  sky fibers depends on the wavelength domain of the observations, the
  dimension of the field-of-view, and the requirement on the quality
  of the sky-subtraction \cite{1992MNRAS.257....1W}. Observations
  need to be previously corrected from the individual response of the
  fibers and scattered light.
\item \textbf{Beam-switching:} Each fiber is alternatively switched
  between the object and a reference sky position (nodding). This
  strategy has the advantage of sampling the sky background with the
  same fibers that those used for observing the targets
  \cite{1998ASPC..152...50W}. The sky and the science signals can
  be subtracted simultaneously. However, this procedure implies to
  dedicate half of the observation time for sky sampling and does not
  account for temporal variations of the sky background
  \cite{1992MNRAS.259..751R}, \cite{2005NewA...11...81L}].
\end{itemize}
However, even with these procedures the subsequent sky subtraction
hardly reaches a quality better than 1\%, i.e., a quality that can be
easily reached with slit spectrographs. Therefore, we have designed a
new method which take advantage of the spatial information provided by
the multi-object mode and of the prior knowledge on the nature of the
sky, and in particular its spatial and temporal variations.

\subsection{Sky variations} 
Knowing how the sky fluctuates in space and time is crucial for the
implementation of an optimal sky subtraction. In this subsection we
briefly describe the main characteristics of the sky variability. 

The sky brightness \footnote[1]{The skylight is made of a mixture of
  radiation produced by several sources such as moonlight, zodiacal
  light, airglow, thermal emission, unresolved background
  astrophysical sources, etc. In this work we have only taken into
  account the dominant source of sky in dark sites and optimal
  observational conditions: the emission of the upper atmosphere named
  airglow.} show significant spatial and temporal variations
\cite{1961paa..book.....C},\cite{1973QB817.R54......}. These
are due to the dynamical nature of the atmosphere. Indeed, the
atmosphere is a complex system in which the main properties at a given
layer -composition, density, temperature - vary with time and space.
The main factor of variation is the evolution of the chemical
composition and density in the mesosphere, involved by harmonic
periods of diurnal tides. Superposed to this smooth diurnal
variations, the airglow emission is also affected by short-scale
variations that are more problematic for spectroscopic observations.
These short-period variations, of the order of few minutes to an hour,
is due to perturbations in density and temperature within the upper
atmosphere. Using FORS2 observations, we found that the sky background
varies across the field (5'$\times$5') by 5-10\%. The intensity of skylines
also vary spatially from 10\% to 20\%. These spatial variations are
smooth with a spatial scale of 1.0', which is coherent with the
passage of gravity waves. In the near-IR (1-1.8$\mu m$),
\cite{1992MNRAS.259..751R}] and \cite{2008ExA....22...87M}
have shown that the fluctuations of the OH emission bands can have
variations of 5-10\% and that the sky background fluctuates with an
amplitude of 15\% (see also \cite{2008MNRAS.386...47E}).

%%%%%%%%%%%%%%%%%%%%%%%%%%%%%%%%%%%%%%%%%%%%%%%%%%%%%%%%%%%%%
\section{A new algorithm for sky extraction} 
\begin{figure}
 \begin{center}
\includegraphics[width=0.95\textwidth]{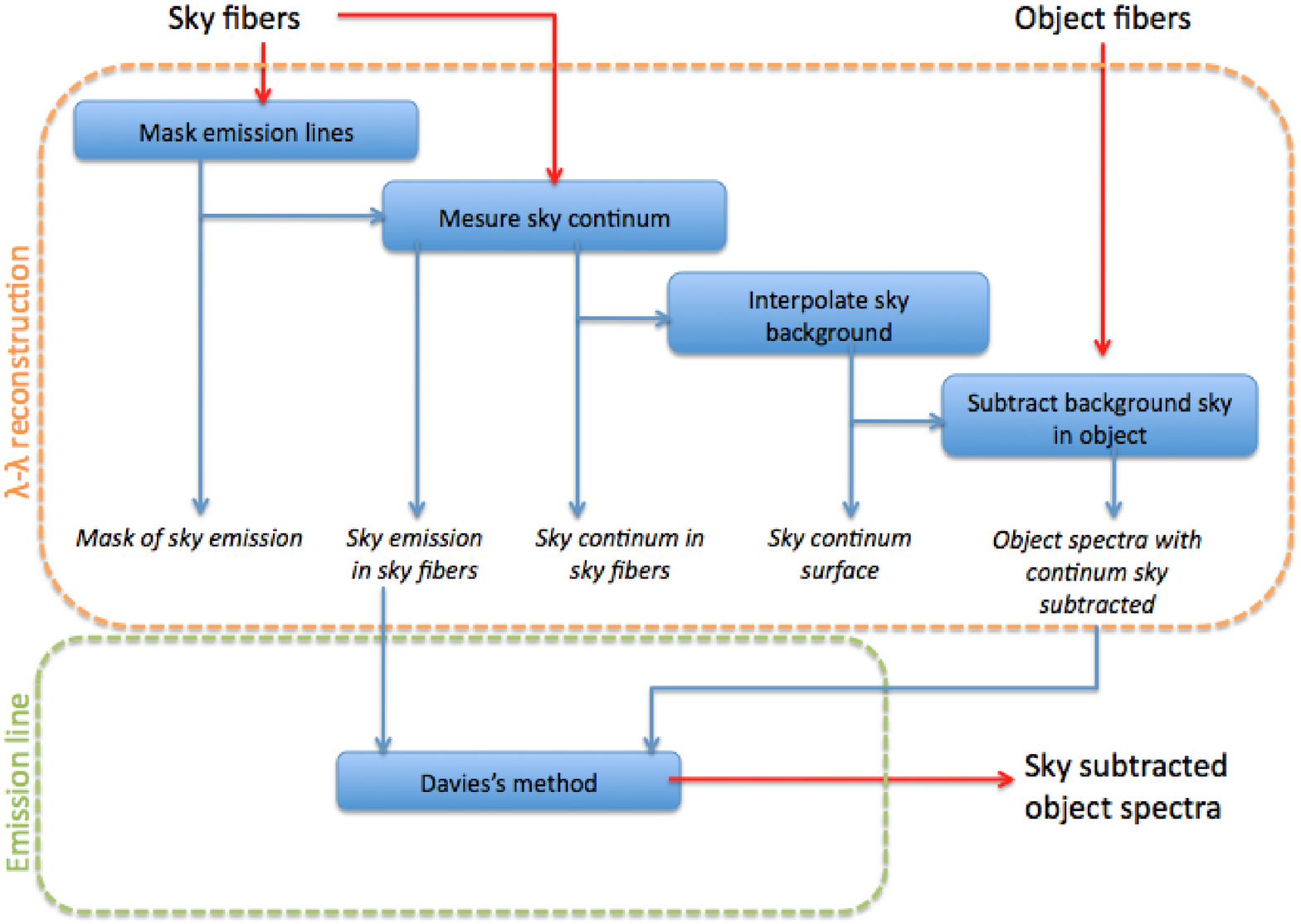}
\includegraphics[width=0.95\textwidth]{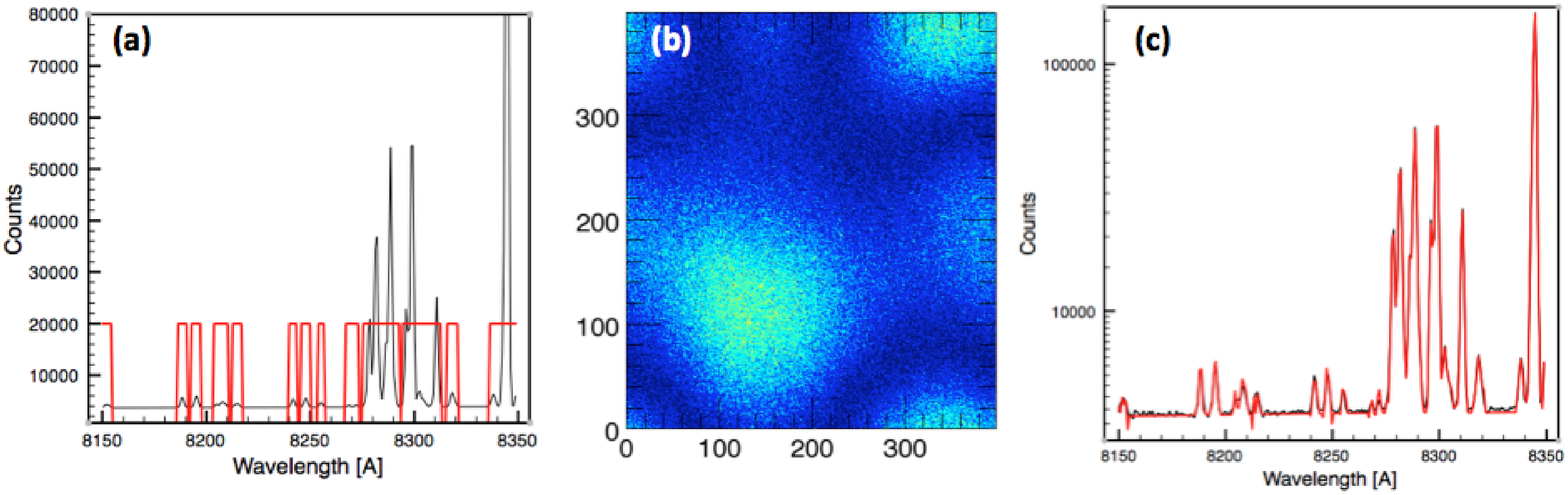}
\caption{Flowchart of the sky subtraction algorithm. Examples are
  shown in the lower panels, with, from left to right: (a) sky
  emission mask (red boxes) produced by sigma-clipping, (b) residuals
  from the subtraction of the sky continuum over a $7'\times7'$ field
  using 120 sky fibres and a 1h exposure time, (c) modeled sky
  background (red line) superimposed to the input sky spectrum (black
  line).}
 \end{center}
\label{Sky_flowchart}
\end{figure}

The characterization of the spatial and temporal variations of the sky
background allows us to optimize the algorithm of sky subtraction. As
sky emission and sky continuum do not vary in phase, one can treat the
two components independently. The algorithm is therefore divided into
two steps: (1) the determination of the sky background using a
$\lambda-\lambda$ surface reconstruction and (2) the extraction of the
emission sky lines using the Davies' method
\cite{2007MNRAS.375.1099D}. A flowchart of the whole sky
extraction algorithm is presented in Fig. \ref{Sky_flowchart}

\subsection{The sky continuum: $\lambda-\lambda$ reconstruction}
The sky continumm is sampled at different spatial positions
distributed over the instrument field of view using dedicated sky
fibers. An interpolation method on an irregular grid is used to
reconstruct a sky continuum surface over the total field of view, as a
function of wavelength in regions free of emission sky lines. The
three main steps are described below:
 \begin{description}
\item[(1) Uncouple the continuum from the emission lines:]

All sky spectra are stacked into a single sky spectrum in which a
sigma-clipping algorithm is used to detect the emission sky lines.
This step produces a sky emission mask, which is used to mask all
emission lines in the individual sky spectra and isolate spectral
regions of pure continuum.

\item[(2) Sky continuum extraction:]

For each sky fibre, emission lines are masked with the sky emission
mask derived from step (1). Regions without emission lines are
filtered spectrally using a wavelet decomposition to improve their
signal-to-noise ratio \cite{1994A&A...288..342S}. The smoothed
continuum is then interpolated over the entire wavelength range in
order to get a pure continuum sky spectrum in each sky fibre.

\item[(3) $\lambda-\lambda$ surface reconstruction:]

For each spectral element of resolution free of emission sky lines, a
sky continuum surface is reconstructed thanks to a triangulation
interpolation method, using the IDL routine GRIDDATA with a natural
neighbor interpolation method.
\end{description}
 
At the end of this step, a $\lambda-\lambda$ sky continuum datacube is
constructed, which summarizes all the constrains on the sky continumm
over the field of view and wavelength bins. The sky continuum at the
position of the sky and object fibres is extracted from this datacube,
and subtracted to the sky and object individual spectra. Sky continuum
subtracted spectra are then sent to the next step of the algorithm.

\subsection{Removing emission sky lines}
We used the method proposed by \cite{2007MNRAS.375.1099D}, which
has been shown to be able to remove the OH lines in the near infrared
(1 to 2.5 $\mu m$) with a good accuracy. This technique takes into
account absolute and relative variations of OH sky line intensities,
as well as variations due to instrumental flexures, which can impact
the wavelength scale. The reader is refereed to
\cite{2007MNRAS.375.1099D} for a detail description of the
algorithm.

\section{Sky subtraction with OPTIMOS-EVE}
\begin{figure}[!h]
\centering
\includegraphics[width=0.80\textwidth]{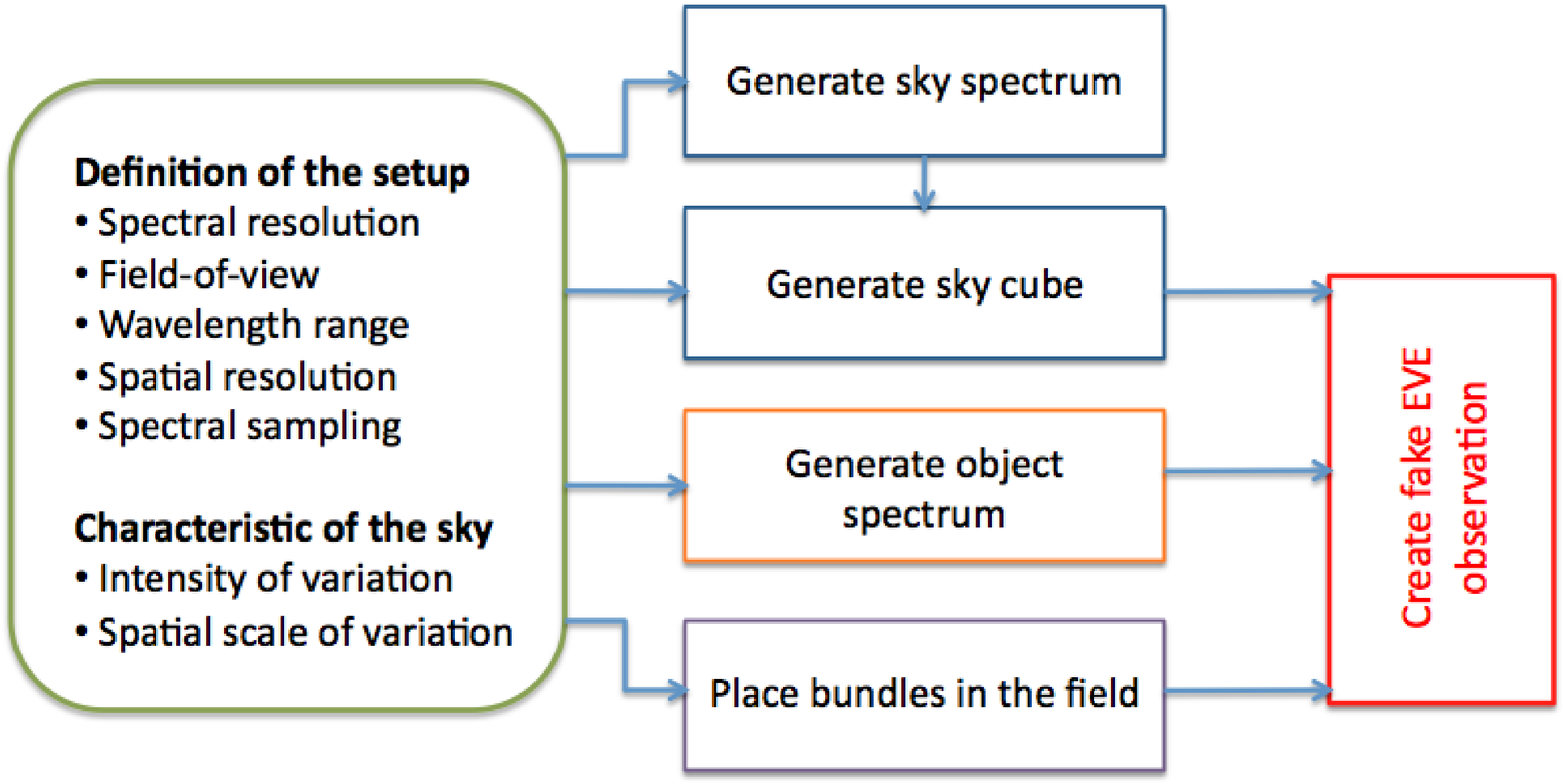}
\caption{Flowchart of the simulator for the mono-fiber and IFU modes.
  The sky spectrum is generated from the combination of a sky
  continuum, given by the Standard ESO ETC sky brightness and
  \cite{2000Msngr.101....2C}, and a library of sky lines from
          \cite{2003A&A...407.1157H}. The object spectrum is a
          high-z galaxy modeled by a constant continuum of magnitude
          $mag_{con}$ and an emission line defined by a Gaussian shape
          centered at the position $\lambda_{emi}$ with a total flux
          of $flux_{emi}$ and a velocity dispersion $\sigma_{emi}$.}
\label{Flowchart_skysimulator}
\end{figure}

In order to test and optimize this algorithm for OPTIMOS-EVE, we have
constructed a simulator for MO and MI EVE observational modes (see
Sect. 1.1). This simulator is composed of a sky generator, which
creates artificial sky datacubes, an object generator, which creates
artifical science spectra, and an interactive routine that allows us
to place IFUs or fibers within the EVE field of view. The parameters
of the simulation have been chosen to be as close as possible of the
EVE observational setups, in terms of spectral resolution, spatial
scale, lambda sampling, and field-of-view. The architecture of the
simulator is described on the flowchart shown in Fig.
\ref{Flowchart_skysimulator}.

\subsection{Accuracy on the sky background extraction}
We have tested whether the designed algorithm can reach the required
accuracy of $<1\%$ on sky extraction required for both MO and MI
observational modes. Hereafter, the quality of the extraction is
defined as the mean residual between the input and the recovered sky
surface as a function of wavelength.

\begin{figure}[h!]
\centering
\includegraphics[width=1.0\textwidth]{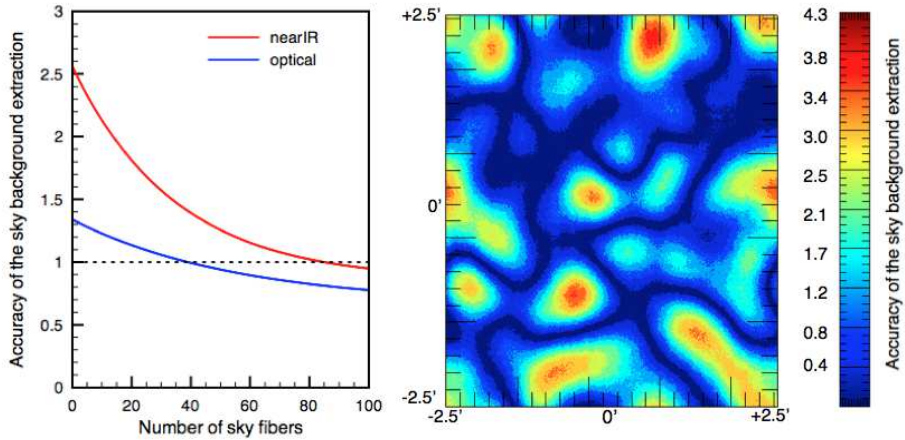}
\caption{{\small Right panel: Accuracy of the sky background
    subtraction as a function of the number of dedicated sky fibers
    for the optical (blue) and near IR (red) wavelength domain. Left
    panel: quality of the extracted sky background in the MO mode at
    visible wavelengths (\%). Simulations were done on a field of
    7x7arcmin with 91 fibers randomly distributed on the sky. 110
    objects fibres have been also distributed in the field of view.
    The quality of the sky background extraction in the object fibre
    is less than 0.9\%.}}
\label{Sky_MO}
\end{figure}

In the MO mode, the quality of the sky continuum surface
reconstruction depends on the number of dedicated sky fibers and how
they are distributed over the field-of-view. We have first tested the
quality of the sky continuum extraction as a function of the number of
dedicated sky fibers. For a given number of sky fibers, we have
simulated 50 observations of one hour exposure with sky fibers
randomly distributed over the field of view, and measured the accuracy
of the sky continuum extraction in a central $5'\times5'$ region.
Figure \ref{Sky_MO} illustrates the results of these Monte-Carlo
simulations using the median value of the extraction quality over the
50 iterations which is plotted as a function of the number of sky
fibers. At visible wavelengths, the sky continuum can be easily
retrieved within a quality of 1\% using 30-40 sky fibers. In the near
IR, due to the more important variations of the sky background (20\%
peak to peak), at least 80 fibers are needed to properly recover the
sky background with the same quality.

In the MI mode, we have simulated the case of an object illuminating
nine central spaxels, which roughly corresponds to median seeing
conditions. The remaining free fibers in the IFU were used as
constraints for interpolating the sky background, in addition to four
dedicated fibers sampling the sky contribution at larger spatial
scales. The MI mode allows us to reach an accuracy on the sky
extraction of 0.3\%. It is thus recommended for observations for which
the sky background has to be recovered with very high accuracy.

\subsection{Application to observations of distant and faint galaxies}
With OPTIMOS-EVE, it is required to observe very faint emission lines,
down to $1\times10^{-19}\,ergs/cm^2/s$, from objects with a continuum
reaching $m_J$= 28 or more. We have verified the feasibility of this
science case in term of sky subtraction, in the MO and MI modes. We
have modelled an emission line (such as the Lyman $\alpha$ line)
falling in a spectral window devoid of strong sky emission lines in
the J-band\footnote[2]{As a reference, the magnitude of the sky
  continuum in this band is 18.0 \cite2000Msngr.101....2C}, i.e., ten
  magnitudes brighter than the object continuum. The test has been
performed using a Monte-simulation of 40 independent observations of
1h exposure (40 different sky cubes) for each mode.

For the mode MO, 82 targets and 120 sky fibers have been distributed
in the EVE field-of-view. Forty independent observations of 1h
exposure have been generated using the EVE simulator. Each individual
exposures have been sky-subtracted using the sky subtraction algorithm
and then combined together with a median min-max rejection algorithm.
Figure \ref{MO_high} summarizes the performance of the MO mode in
recovering a distant galaxy with an AB magnitude of 28, an emission
line flux of $1\times10^{-19}\,ergs/cm^2/s$, and a
$FWHM_{obs}$=150km/s, in 40 hr of integration time. The emission line
is recovered with a mean S/N ratio of 8.

 \begin{figure}[h!]
\centering \includegraphics[width=1.0\textwidth]{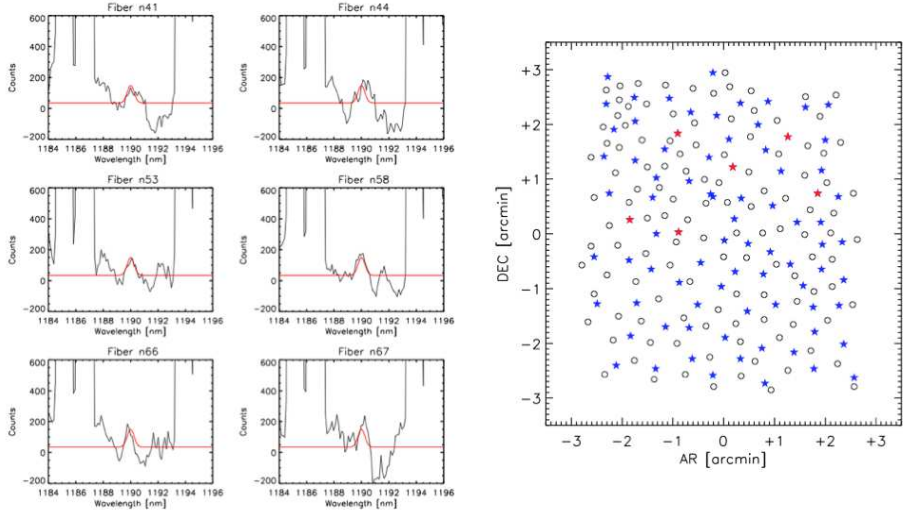}
\caption{{\small Simulation of 82 galaxies at z=8.8 in the J-band
    observed during 40 hours in the MO mode with ground-layer adaptive
    optics (GLAO). All galaxies have a continuum in the J-band of
    m(AB)=28 and the Lyman alpha line has an integrated flux of
    $1\times10^{-19}\,ergs/cm^2/s$ and FWHM=150km/s. Left panels: the
    Lyman alpha line detected after sky background subtraction in
    different fibers (black line). The input spectrum is shown in red.
    The S/N of the detection depends on the quality of the sky
    background subtraction within each fiber. Emission sky lines have
    not been subtracted since such distant object emission lines are
    usually targeted in regions free of OH sky lines to boost the
    observational efficiency. Two sets of sky lines are visible at
    1184-1187nm and 1193-1195nm. Right panel: EVE field-of-view and
    observational setup. The sky has been sampled with 120 fibers
    (half of the total number of available MO bundles; see open
    circles). The 82 simulated galaxies are shown as blue stars, while
    the red stars correspond to the fibers for which the measured
    spectra are shown in the left panels.}}
\label{MO_high}
\end{figure}

\begin{figure}[h!]
\centering
\includegraphics[width=1.0\textwidth]{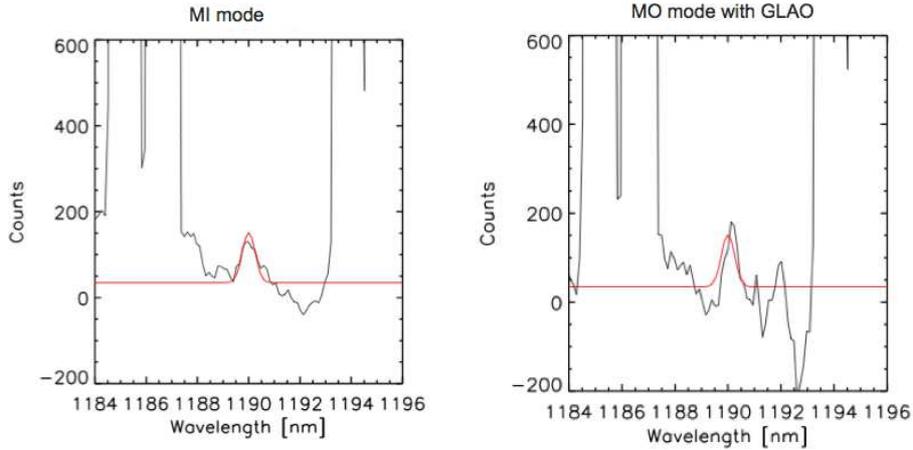}
\caption{{\small Simulation of a z=8.8 galaxy observed in the J band
    during 40 hours in the MI (left panels) and MO+GLAO (right panel)
    modes. The galaxy has a J-band continuum of m(AB)=28 and the Lyman
    alpha line has an integrated flux of
    $1\times10^{-19}\,ergs/cm^2/s$ with FWHM=150km/s. Left panel: the
    Lyman alpha line is detected in the stacked central IFU spaxels
    after sky background subtraction (black line). The target
    illuminates nine central spaxels that have been summed up. The
    remaining spaxels are used for sky sampling. Right panel:
    simulation of observations in the MO mode of the same Lyman alpha
    line. The observational setup is the same that the one described
    in Fig. \label{MO_high}.}}
\label{Compare_MOvsMI}
\end{figure}

In MI mode, the galaxy illuminates nine spaxels of the IFU. The other
spaxels of the IFU are used for the determination of the sky, in
complement of the four surrounding sky fibers. As the MO mode, 40
independent observations of one hour exposure have been simulated. The
sky has been extracted in each individual exposure, using an optimal
extraction to subtract the sky contribution. The central IFU pixels
have been stacked together, and then sky-subtracted. The MI mode
allows us to reach an S/N of 5 and 16 for faint emission lines
($1\times10^{-19}\,ergs/cm^2/s$) and an underlying continuum of 30 mag
and 28 mag, respectively. Figure \ref{Compare_MOvsMI} shows the
detected emission line after sky background subtraction (exposure time
of 40 hours) in the MI and MO modes. In the MI mode, the sky
background is better subtracted and the line is detected with a better
S/N than with the MO mode.

\section{Conclusion}
We have presented a new algorithm for sky subtraction dedicated to
fibre-fed spectrographs, such as E-ELT/OPTIMOS-EVE. Spectroscopy of
faint objects with fibre-fed instruments is often thought to be
limited by the inhability of such instruments to measure the sky
contribution close enough to science targets. Using Monte-Carlo
simulations, we have demonstrated that this new algorithm can
nevertheless reach accuracies similar to slit spectrographs. This is
due to a careful reconstruction of a sky continuum surface as a
function of wavelength using dedicated sky fibres distributed over the
whole instrument field-of-view. This makes it possible to interpolate
the sky contribution at the location of the science channels with good
accuracy. Objects as faint as
$f_{Ly_{\alpha}}$=$1\times10^{-19}\,ergs/cm^2/s$ and $m_{cont} =30$,
which fairly represent z=8.8 Lyman-alpha emitters, should be
detectable by the IFU mode of OPTIMOS-EVE within 40 hours of
integration time.

%%%%%Sometimes it is necessary to precede the double slash 
%%%%%by \verb|\protect| to get the desired result, 
%%%%%for example, in article titles.

%%%%%%%%%%%%%%%%%%%%%%%%%%%%%%%%%%%%%%%%%%%%%%%%%%%%%%%%%%%%%
%\acknowledgments %>>>> equivalent to \section*{ACKNOWLEDGMENTS} 

%This unnumbered section is used to identify those who have aided the authors in understanding or accomplishing the work presented and to acknowledge sources of funding.  

%%%%%%%%%%%%%%%%%%%%%%%%%%%%%%%%%%%%%%%%%%%%%%%%%%%%%%%%%%%%%
%%%%% References %%%%%

\bibliography{OPTIMOS_EVE.bib}   %>>>> bibliography data in report.bib
\bibliographystyle{plain}   %>>>> makes bibtex use spiebib.bst

\end{document}